\documentclass[aps,pre,twocolumn,groupedaddress,floatfix,amsmath]{revtex4}
\usepackage{graphicx}
\usepackage{amsmath,amssymb}
\usepackage[utf8]{inputenc}
\usepackage{psfrag}
\usepackage{bm}

\begin{document}

\title{New ordered phases in a class of generalized XY models}

\author{Fábio C. Poderoso}
\affiliation{Instituto de F\'\i sica, Universidade Federal do 
Rio Grande do Sul, CP 15051, 91501-970 Porto Alegre RS, Brazil} 
\author{Jeferson J. Arenzon}
\affiliation{Instituto de F\'\i sica, Universidade Federal do 
Rio Grande do Sul, CP 15051, 91501-970 Porto Alegre RS, Brazil} 
\author{Yan Levin}
\affiliation{Instituto de F\'\i sica, Universidade Federal do 
Rio Grande do Sul, CP 15051, 91501-970 Porto Alegre RS, Brazil}

\date{\today}

\begin{abstract}
It is well known that the 2d XY model exhibits an unusual infinite order phase transition  belonging to the 
Kosterlitz-Thouless (KT) universality class. Introduction of a nematic coupling into the XY Hamiltonian  
leads to an additional phase transition in the Ising universality class~\cite{LeGr85}.  
In this paper, using a combination of extensive Monte Carlo simulations and finite size scaling, we  show  
that the higher order harmonics lead to a qualitatively different phase diagram, with additional ordered phases
originating from the competition between the ferromagnetic and pseudo-nematic couplings. 
The new phase transitions belong to the 2d Potts, Ising, 
or KT universality classes. 
\end{abstract}

\maketitle


The low temperature behavior of two dimensional (2d) systems with continuous symmetries is controlled by 
topological defects, such as vortices and domain walls.  Although massless Goldstone excitations,
such as spin waves, destroy the long-range order of these systems, a pseudo-long-range order with algebraically 
decaying correlation functions still remains possible.  At low temperatures the 
topological defects which undermine the pseudo-long-range order are all paired up, while above the critical 
temperature these defects unbind, leading to exponentially decaying correlation functions
and a loss of the pseudo-long-range order. A classical example of such system is the
XY model. At low temperature, the topological defects, in the form of integer valued vortices, are all joined 
in vortex-antivortex pairs, resulting in algebraically decaying spin-spin
correlation functions.  Above the Kosterlitz-Thouless (KT) critical temperature~\cite{KoTh73,Kosterlitz74},
these pairs unbind and the 
correlation functions decay exponentially. 

Unlike in 3d, for 2d systems the
arguments based purely on symmetry considerations are not sufficient to fix the 
universality class of possible phase transitions~\cite{Baxter71,DoScSw84,Vinck07,Levin07}, and
even in 3d a second order phase transition can be preempted by 
a first order one~\cite{EnSh02}.  Violations of strong universality
are even more common in 2d.
Thus, it is possible for  systems with the same underlying symmetries and the
same coarse-grained Landau-Ginzburg-Wilson Hamiltonian not to belong to the same 
universality class. It is, therefore,
interesting to ask what phase transitions are possible for 2d Hamiltonians
invariant under the transformation $\theta \rightarrow \theta + 2 \pi$.  
In this paper, we will study using extensive Monte Carlo simulations and finite 
size scaling (FSS) analysis, a large class of generalized XY models which,
while preserving the same  $\theta \rightarrow \theta + 2 \pi$  symmetry, have very complex phase diagrams,
with phase transitions belonging to the  Ising and Potts universality classes, 
in addition to the usual KT phase transition.  
In some of these models, 
transitions can be understood in terms of new topological defects, such as fractional vortices and 
domain walls~\cite{Korshunov85,LeGr85}. Apart from the fundamental considerations regarding the connection 
between symmetry and universality, our purpose is to describe new, previously unnoticed, ordered phases which occur in 2d systems
with continuous symmetry. 

The model considered here has a mixture of ferromagnetic and nematic-like interactions,
\begin{equation}
\mbox{H} = - \sum_{\langle i j \rangle} [\Delta\cos(\theta_i - \theta_j) +  (1 - \Delta) \cos
(q\theta_i - q\theta_j)],
\label{eq.hamiltonian}
\end{equation}
where the sum is over the nearest neighbors spins on a square lattice, $0\leq\Delta\leq 1$, and 
$q$ is a positive integer. The first term is the usual ferromagnetic coupling (XY model), while the second one 
favors  the adjacent 
spins to have a phase difference of $2k\pi/q$, where $k \le q$ is an integer. Independent of the value of $\Delta$, the 
Hamiltonian (\ref{eq.hamiltonian}) has the symmetry of the pure XY model, recovered when $\Delta=1$,
and is invariant under rotations $\theta_j \rightarrow \theta_j + 2 \pi$. For $\Delta=0$, we have a purely nematic-like 
Hamiltonian, which is also invariant under the transformation
$\theta_j \rightarrow \theta_j + 2\pi/q$.  It is easy to show  that in this case there will also be a
KT phase transition at exactly the same critical temperature as in the pure XY model.  The low temperature 
phase for $\Delta=0$ will, therefore, have a
pseudo-long-range nematic-like order.  An interesting question concerns the thermodynamics of the model described
by Eq.~(\ref{eq.hamiltonian}) for $0<\Delta<1$, where both terms compete.

\begin{figure*}[tb]
\includegraphics[width=8.cm]{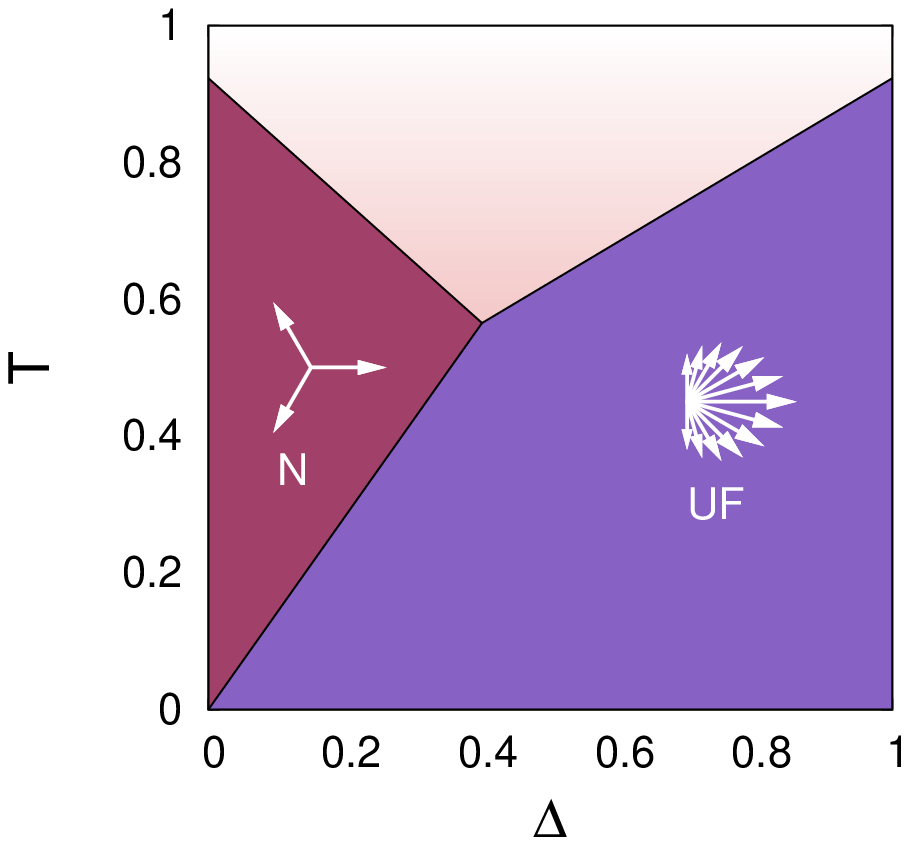}
\includegraphics[width=8.cm]{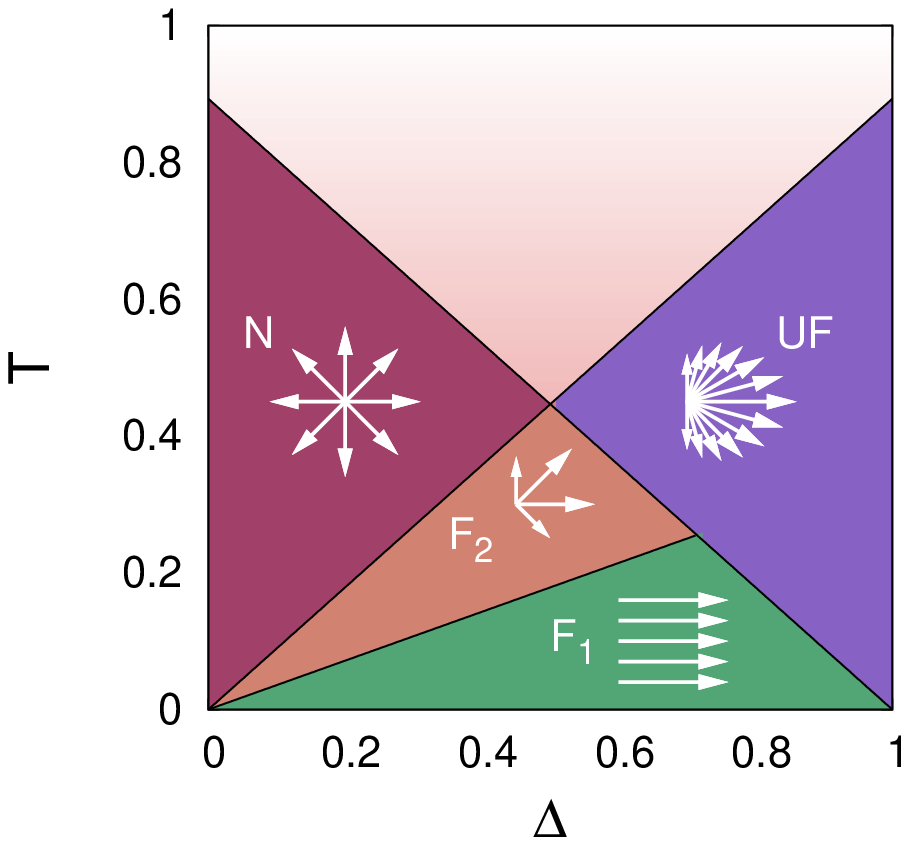}

\hspace{1.6cm} (a) $q=3$\hspace{6.5cm}(b) $q=8$
\caption{a) Schematic phase diagram for $q = 3$. At both extremes $\Delta=0$ and $1$, the order-disorder 
transition temperature is $T_c=0.893$. There are two low temperature phases with a pseudo-long range order:
generalized-nematic (three preferred spin orientations) and ferromagnetic  (broken reflection symmetry). 
 Notice the multicritical point around $\Delta_{mc}\simeq 0.4$. For $\Delta<\Delta_{mc}$, there are
two transitions: paramagnetic to nematic, in the KT universality class; and nematic to ferromagnetic, in the $q=3$ Potts 
universality class. b) Schematic phase diagram for $q = 8$. Besides the paramagnetic, the usual ferromagnetic, and the nematic phases,
there are two new ferromagnetic phases (F$_1$ and F$_2$) in which spins have half-plane preferred orientations.}
\label{fig.diagramq3}
\end{figure*}

For $q=2$, the Hamiltonian, eq.~(\ref{eq.hamiltonian}),  has been studied by a number of
authors~\cite{Korshunov85,LeGr85,SlZi88,CaCh89,PaOnNaHa08,KoHl10} and the presence of the  second term 
leads to metastable states in which spins have antiparallel orientation.  
The model has new excitations not present in the $\Delta=1$ case: half-integer vortices connected by strings (domain
walls)~\cite{LeGr85}, 
across which spins are anti-paralelly aligned. At low temperatures, half-integer vortices are bound in pairs of
integer vorticity, resulting in a pseudo-long-range ferromagnetic order.  
If $\Delta<\Delta_{mc}$, as the temperature is raised, the string tension between  half-integer vortices
vanishes and the system melts into a nematic phase.  On further increase of temperature,
the half-integer vortex-antivortex pairs unbind and the system enters a completely disordered  paramagnetic phase.  
As expected, the transition between the nematic and the paramagnetic phases belongs to the KT universality class~\cite{BaSt07a}.
Surprisingly, the transition between the two pseudo-long-range 
ordered phases --- nematic and ferromagnetic  --- is found to be in the Ising 
universality class~\cite{LeGr85}.  
This behavior has been verified with simulations on the square and triangular 
lattices~\cite{CaCh89,PaOnNaHa08}. 
On the latter, the geometric frustration introduces also a tiny
chiral phase above the KT line,  but otherwise the phase diagram retains the same 
topology as on the square lattice~\cite{PaOnNaHa08}. 
A related model, with a similar phase diagram, was also studied in Ref.~\cite{GeSe09}.
An interesting question is whether, for $q>2$, the topology of the phase diagram remains unchanged. 
To answer this, we have explored using extensive Monte Carlo simulations 
the equilibrium phase diagram of the  Hamiltonian~(\ref{eq.hamiltonian})
with $q$ ranging from 2 to 10.
We find that for $q=2,3,4$ the topology of the phase diagram remains the same as for $q=2$, 
however, the transition between the pseudo-ferromagnetic and
pseudo-nematic phases belongs either to the 3-state Potts ($q=3$) or to the 2d Ising ($q=2,4$)
universality class, see Fig.~\ref{fig.diagramq3}a.  For $q \ge 5$, the
topology of the phase diagram changes completely and new phases with a 
pseudo-long range order come into existence.




The simulations were performed on a square lattice of linear size $L$ and 
periodic boundary conditions. Both Metropolis single-flip and the Wolff algorithm~\cite{Wolff89}
were used. In accordance with the symmetry of the Hamiltonian, the possible order parameters are
$
m_k=L^{-2}\left| \sum_i \exp(ik\theta_i)\right|
$ 
where $k = 1,\ldots, q$. The corresponding generalized susceptibilities are  $\chi_k = \beta L^2 (\langle m_k^2 \rangle-
\langle m_k\rangle^2)$ and the Binder cumulants are 
$U_k = 1 - \langle m_k^4 \rangle/3 \langle m_k^2 \rangle^2$~\cite{Loison99,Hasenbusch08}. 
If the transition is not KT, the usual FSS can be used to get the critical exponents $\beta, \gamma,$ and $\nu$:
$m = L^{-\beta/\nu} f(t L^{1/\nu})$  and $\chi = L^{\gamma/\nu} g(t L^{1/\nu})$, where $m$ is the 
order parameter and $\chi$ its susceptibility,  $f$ and $g$ are the scaling 
functions,  and $t=T/T_c-1$ is the reduced temperature. 
This FSS, however,  is not valid at the KT transition
for which all of the low temperature phase is critical and the correlation length and 
the susceptibility are infinite~\cite{Kosterlitz74,ToCh79}. Nevertheless, it is possible to show that at 
the  KT transition and in the low-temperature phase, the critical exponent {\it ratios} are well defined and the
order parameter and the generalized susceptibility scale with the size of the system as $m \propto L^{-\beta/\nu}$ and 
$\chi \propto L^{\gamma/\nu}$.
Exactly at the transition, $\beta/\nu=1/8$ and $\gamma/\nu=7/4$, which are the same ratios as for the 2d Ising model.  However,
what distinguishes the KT transition from the Ising one, is the behavior of the order parameter and the susceptibility in the low
temperature phase where they also exhibit FSS, but with non-universal critical exponents. 
Recall that for normal second order phase transition, FSS exists only at the critical point.  This difference can be used
to distinguish the KT transition from the Ising one.

\begin{figure}[tb]
\vspace{4mm}
\includegraphics[scale=0.28,angle=270]{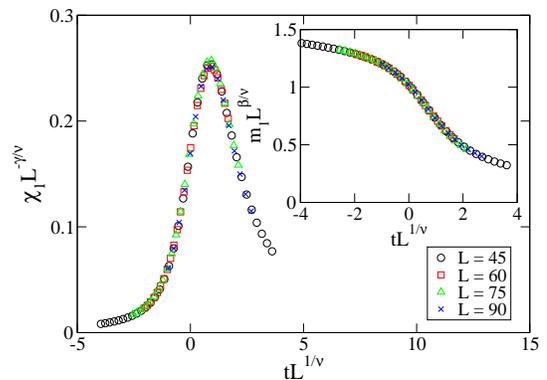}
\caption{Ferromagnetic-nematic transition for $q=3$. FSS analysis of 
the susceptibility $\chi_1$ and the order parameter $m_1$ (inset)
for $\Delta = 0.25$ and different sizes $L$. In both cases the data collapse is excellent using
the 3-state Potts exponents. The critical temperature at this point is  $T_c\simeq $ 0.365.} 
\label{fig.fssq3}
\end{figure}

Figure \ref{fig.diagramq3}a shows the schematic phase diagram for $q=3$, which is topologically identical to the  $q=2$ case. 
At high temperatures, the equilibrium state is the disordered paramagnetic (P). As the temperature is lowered, 
the system enters either the usual ferromagnetic (UF) or the generalized-nematic phase (N), depending on the value of $\Delta$. 
Both of these order-disorder transitions belong to the KT universality class. Up to the multicritical point located 
at $\Delta_{mc}\simeq 0.4$, there is a line of critical points separating the generalized-nematic from the UF phase. 
The order parameter $m_1$ is used to distinguish between the generalized-nematic and the ferromagnetic phases:
$m_1\simeq 0$ in the nematic phase and is $\approx 1$ in the ferromagnetic phase. 
Using FSS, and also the Binder cumulant, 
we find that the critical points along this line belong to the 3-state Potts universality class. 
Fig.~\ref{fig.fssq3} shows the data collapse of $m_1$ (inset) and the corresponding susceptibility
$\chi_1$, for a critical point with $\Delta=0.25$. The collapse is excellent using the critical exponents of the 2d, 3-state Potts
model~\cite{Wu82}:  $\beta=1/9$, $\gamma=13/9$ and $\nu=5/6$.


\begin{figure}[tb]
\vspace{5mm}
\includegraphics[scale=0.5,angle=270]{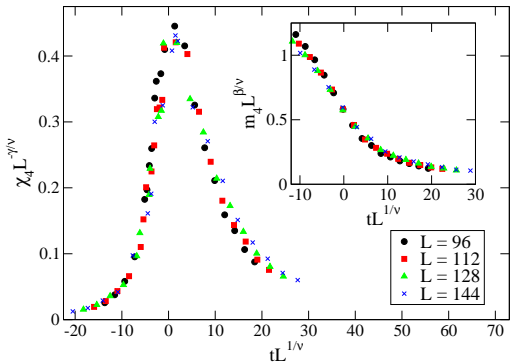}
\caption{Data collapse for the susceptibility $\chi_4$ and the order parameter $m_4$ (inset) for $q = 8$  at  $\Delta$ = 0.15
as the system crosses the border between F$_1$ and  F$_2$ phases.
The collapse was obtained using the critical exponents of the $2d$ Ising model,
$\gamma = 7/4$, $\nu = 1$, $\beta = 1/8$ and $T_c\simeq 0.058$.} 
\label{fig.F-F1}
\end{figure}

\begin{figure}[tb]
\includegraphics[width=8cm]{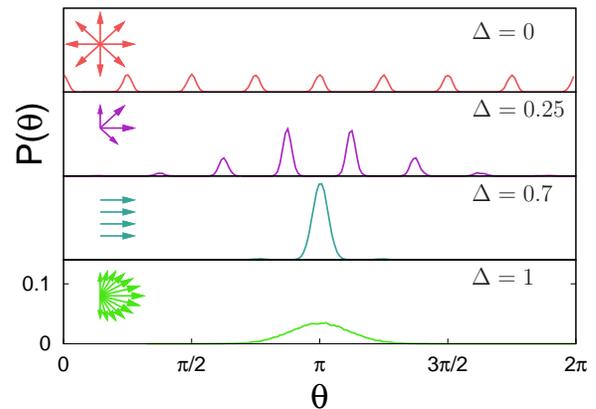}
\caption{Angle distribution for the four phases along the
fixed temperature line $T=0.16$ of the $q=8$ phase diagram. All graphs have the same vertical scale.} 
\label{fig.distributions}
\end{figure}

The same topology of the phase diagram persists for $q=4$, except that the transition from the  generalized-nematic to the 
ferromagnetic phase is once again in the universality class of the Ising model~\cite{Betts64,BaMi10}.  For $q\geq 5$, however, the topology
changes dramatically.  Two new phases emerge from  the $T=0$, $\Delta=0$ fixed point, see   
Fig.~\ref{fig.diagramq3}b. 
The new phases are pseudo-ferromagnetic and have a broken reflection symmetry.  We shall denote these 
phases as F$_1$ and F$_2$.  The low-temperature phase F$_1$ has only one preferred spin orientation, while
in the phase  F$_2$ there are four preferred spin orientations with different weights, 
see  Figs.~\ref{fig.diagramq3}b and \ref{fig.distributions}. 
Fig.~\ref{fig.distributions} exhibits histograms of spin orientation in the low temperature phases along the line $T=0.16$ 
together with a pictorial representation of the possible spin orientations for the $q=8$ model.  
All the ferromagnetic phases have a quasi-long-range order and a broken reflection symmetry.
For F$_2$, the distribution function has four significant peaks, while for  F$_1$ there is only a single narrow peak. 
The UF phase of the XY model has a broad continuous distribution of spin orientations.
The nematic phase is characterized by eight congruent discrete 
spin orientations, separated
by $\pi/4$.



\begin{figure}[htb]
\vspace{5mm}
\includegraphics[scale=0.28,angle=270]{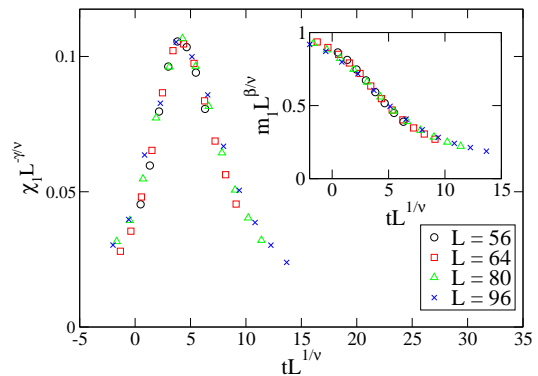}
\caption{Data collapse for the susceptibility $\chi_1$ and the order parameter $m_1$ (inset) for $q = 8$ at $\Delta$ = 0.35
as the system crosses the border between the generalized-nematic and F$_2$ phases at $T_c\simeq 0.34$.
The collapse was obtained using the 2d Ising model critical exponents,
$\gamma = 7/4$, $\nu = 1$ and $\beta = 1/8$.} 
\label{fig.N-F1}
\end{figure}
    
\begin{figure}[tb]
\includegraphics[width=8cm]{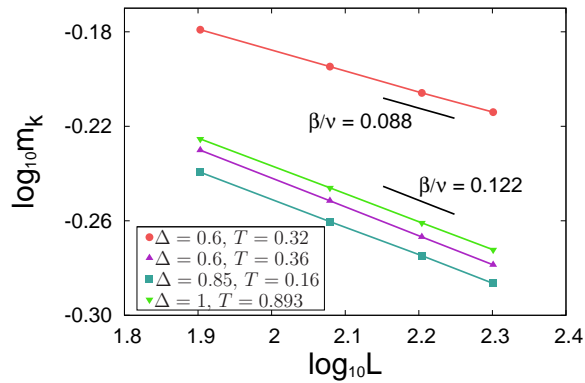}
\caption{Scaling of the order parameter at the KT phase boundaries (three bottommost lines) and inside the critical
phase (topmost line) for $q = 8$. Notice that along the KT transition the exponent $\beta/\nu$ has the Ising value ($1/8$),
while inside the critical phase the exponent is non universal.} 
\label{fig.m_kt}
\end{figure}

Besides the order-disorder KT transitions, several new order-order transition lines appear in the $q \ge 5$ phase diagrams, 
Fig.~\ref{fig.diagramq3}b. The transition from 
the F$_1$ to F$_2$ phase is well described by the $m_4$ order parameter, which is close to
zero in the later phase. The data collapse for several lattice sizes is shown in Fig.~\ref{fig.F-F1} along 
with the data collapse for the susceptibility $\chi_4$. Both collapses were obtained using the set of critical 
exponents of 2d Ising model: $\gamma = 7/4$, $\nu = 1$ and $\beta = 1/8$. 
The transition from F$_2$ to nematic is also continuous 
and is also in 2d Ising universality class, as can be seen in Fig.~\ref{fig.N-F1} in which the data collapse of
the order parameter and the corresponding susceptibility are shown. For this transition, the relevant 
order parameter is  $m_1$. 
Finally, the transition from F$_1$ to UF and the transition from F$_2$ to UF are both well described by 
the $m_8$ order parameter.  The two transitions are found to belong to the KT universality class with the 
critical exponent ratios,  $\gamma$/$\nu$ = 7/4 and
$\beta$/$\nu$ = 1/8.  As an example, Fig.~\ref{fig.m_kt} shows that F$_2$ is critical with respect to $m_8$ order parameter
and has a non-universal FSS characteristic of a low-temperature KT phase, top-most curve of Fig.~\ref{fig.m_kt}.

To conclude, we have studied, using extensive numerical simulations,  phase diagrams of a class of generalized XY models described by
eq.~(\ref{eq.hamiltonian}).  Previous results show that for $q=2$  besides the usual KT transition,
there is also a nematic to ferromagnetic transition in the Ising universality class. We find that 
for $q=3$ and 4 the topology of the phase diagram remains unchanged, but the ferromagnetic-generalized-nematic transition
belongs to the universality class of $3$-state Potts and the Ising model, respectively.  
For $q=5$, the topology of the phase diagram
changes dramatically and two new ferromagnetic phases appear.  After this, up to $q=10$, the maximum value explored in this
work, the topology of the phase diagram remains unchanged. It is very curious that although the systems studied here are 
invariant under the continuous global symmetry $\theta_i\rightarrow \theta_i+\alpha$ (for all spins
simultaneously), for arbitrary
$\alpha$, the sequence of the 
phase transitions between the pseudo-ordered phases follow 
the one observed for the discrete clock models: q=2 Ising; q=3 Potts;
q=4 Ising; for q=5 a bifurcation and a new phase transition 
appears~\cite{ElPeSh79}. If this analogy persists, we expect 
that there should be a critical value of $q$ above which the phase 
transition between F$_1$ and F$_2$ becomes KT~\cite{LaPfWe06}.
This will be explored in a future work.
The present paper also shows a significant lack of
universality of 2d systems: while all the Hamiltonians studied in this paper have the same underlying
symmetry, $\theta \rightarrow \theta + 2\pi$, the transitions between the different phases 
belong to a variety of universality classes.  Furthermore, since the Hamiltonians
discussed in this paper can be thought of as the leading orders in a
Fourier expansion of a general microscopic spin-spin interaction $V(\theta_i-\theta_j$), 
the work raises a troubling question: How much can we really deduce about the
thermodynamics of 2d systems from the form of their coarse-grained Landau-Ginzburg-Wilson (LGW) Hamiltonian?  It is
clear that the symmetry arguments alone are not sufficient to determine the phase diagram of these systems, and one
needs to have a detailed knowledge of the microscopic interactions~\cite{Levin07}.

{\it Acknowledgments.} We thank D. Stariolo for discussions and the Brazilian
agencies CNPq and Fapergs for partial support. JJA acknowledges support
from the INCT-SC while
YL acknowledges support from INCT-FCx, and US-AFOSR/FA9550-09-1-0283.

\bibliographystyle{apsrev}
\bibliography{poderoso}


\end{document}